\documentclass[aps,preprint]{revtex4}

\usepackage{amssymb}
\usepackage{amsmath}
\usepackage{epsfig}
\usepackage{epstopdf}
\usepackage{bm}
\usepackage{graphicx,epsfig}
\usepackage{mathrsfs}
\usepackage{dcolumn}
\usepackage{color}
\usepackage{natbib}
\usepackage{CJK}
\def\be{\begin{equation}}
\def\ee{\end{equation}}
\def\bea{\begin{eqnarray}}
\def\eea{\end{eqnarray}}

\allowdisplaybreaks[2]

\begin{document}

\title{Tunable Topologically-protected Super- and Subradiant Boundary States in One-Dimensional Atomic Arrays}
\author{Anwei Zhang$^{1}$, Xianfeng Chen$^{1,2}$, Vladislav V. Yakovlev$^{3}$, and Luqi Yuan$^{1,2,*}$}
\affiliation{$^1$School of Physics and Astronomy, Shanghai Jiao Tong University, Shanghai 200240, China \\
$^2$State Key Laboratory of Advanced Optical Communication Systems and Networks, Shanghai Jiao Tong University, Shanghai 200240, China \\
$^3$Texas A$\&$M University, College Station, Texas 77843, USA \\
$^*$Corresponding author: yuanluqi@sjtu.edu.cn}



\begin{abstract}
Single-photon super- and subradiance are important for the quantum memory and quantum
information. We investigate one-dimensional atomic arrays under the
spatially periodic magnetic field with a tunable phase, which
provides a distinctive physics aspect of revealing exotic
two-dimensional topological phenomena with a synthetic dimension.
A butterfly-like nontrivial bandstructure associated with the
non-Hermitian physics involving strong long-range interactions has
been discovered. It leads to pairs of topologically-protected edge
states, which exhibit the robust super- or subradiance behavior,
localized at the boundaries of the atomic arrays. This work opens
an avenue of exploring  an interacting quantum optical platform
with synthetic dimensions pointing to potential implications for
quantum sensing as well as the super-resolution imaging.
\end{abstract}

\maketitle







The atomic arrays refer to an ensemble of atoms where the interaction of individual atoms and the photon takes place \cite{ww}. The light-atom coupling in atomic arrays exhibits fundamental physical phenomena including facilitating the long-range coherent interactions and promoting the collective radiative loss \cite{di1}. Recent advances in assembling highly ordered one-dimensional ($1$D) and two-dimensional ($2$D) atomic arrays  provide unique platforms for exploring the  strong light-matter interaction in quantum optics \cite{2,3,6}. The strong interference in the emitted optical field leads to remarkable optical properties such as the subradiant state \cite{s1,s3,s51}, a high reflection of radiation \cite{s6,s8,s0}, the efficient storage and retrieval for quantum memory \cite{m1}, and topologically-protected edge states \cite{l2,l4}, which therefore shows important applications towards quantum information processing, quantum metrology, and nonlinear optics \cite{k1}.

Topological physics is of fundamental importance where physical characteristics are robust against microscopic variation of system details  \cite{ef1,ef2,ef3}. Topological phenomena can be explored by engineering the Hamiltonian of an atomic or optical system \cite{tp3,gg4,gg5}. Such approach shows a great potential towards quantum simulation of the topological matter. The topology in atomic or optical systems provides a novel fundamental way of manipulating quantum states of the light, such as robust photon transport in photonic systems \cite{h1,h2,h3,h4} and non-reciprocal transport in hot atomic gas \cite{dw1,dw2}. Recently, it has been shown that atomic arrays hold a promise for studying topological quantum optics, where the inherent nonlinearity brings a natural way to explore the interacting topological physics \cite{l2,l4,l3,g3,gg3}.

Robust single-photon super- and subradiant states
hold a significant promise for applications related to quantum storage
and quantum information. In this paper, we investigate $1$D
atomic arrays subjected to a spatially periodic magnetic field.
The spatial phase of the magnetic field is an external parameter,
and can be used to map one momentum dimension in a $2$D system
\cite{o1,o02}. Therefore, $1$D atomic arrays with the synthetic
momentum dimension manifests important topological features
associated with $2$D systems. Systems with synthetic dimensions
simplify experimental design and enable capabilities of
manipulating atomic quantum states or photons along the synthetic
dimension \cite{o1,im3,im6,im2,o02,im8}. By changing the
periodicity of the magnetic field, we show that the $1$D atomic
arrays exhibit a butterfly-like spectrum, which
has not been discussed in the 2D atomic arrays under a uniform
magnetic field \cite{l2,l3}. Such spectrum, associated with the
open quantum optical system involving long-range interactions
along the synthetic dimension, exhibits features, which are dramatically distinct from
the spectrum in the 1D photonic model \cite{new1,o2}. For a finite
1D atomic arrays, the system supports pairs of
topologically-protected boundary states with opposite circularly
polarizations, which are found to exhibit super- or subradiance
depending on the magnetic field distribution and atomic excitation
frequency. The topologically-protected subradiant state localized
at the boundary of atomic arrays provides a potential application
towards robust quantum storage under the topological protection.
The results discussed here show
 a unique route towards exploring the strong long-range interacting topological physics in quantum optical system with the synthetic dimension.

 \begin{figure}[htbp]
\centering
\includegraphics[width=0.5\textwidth ]{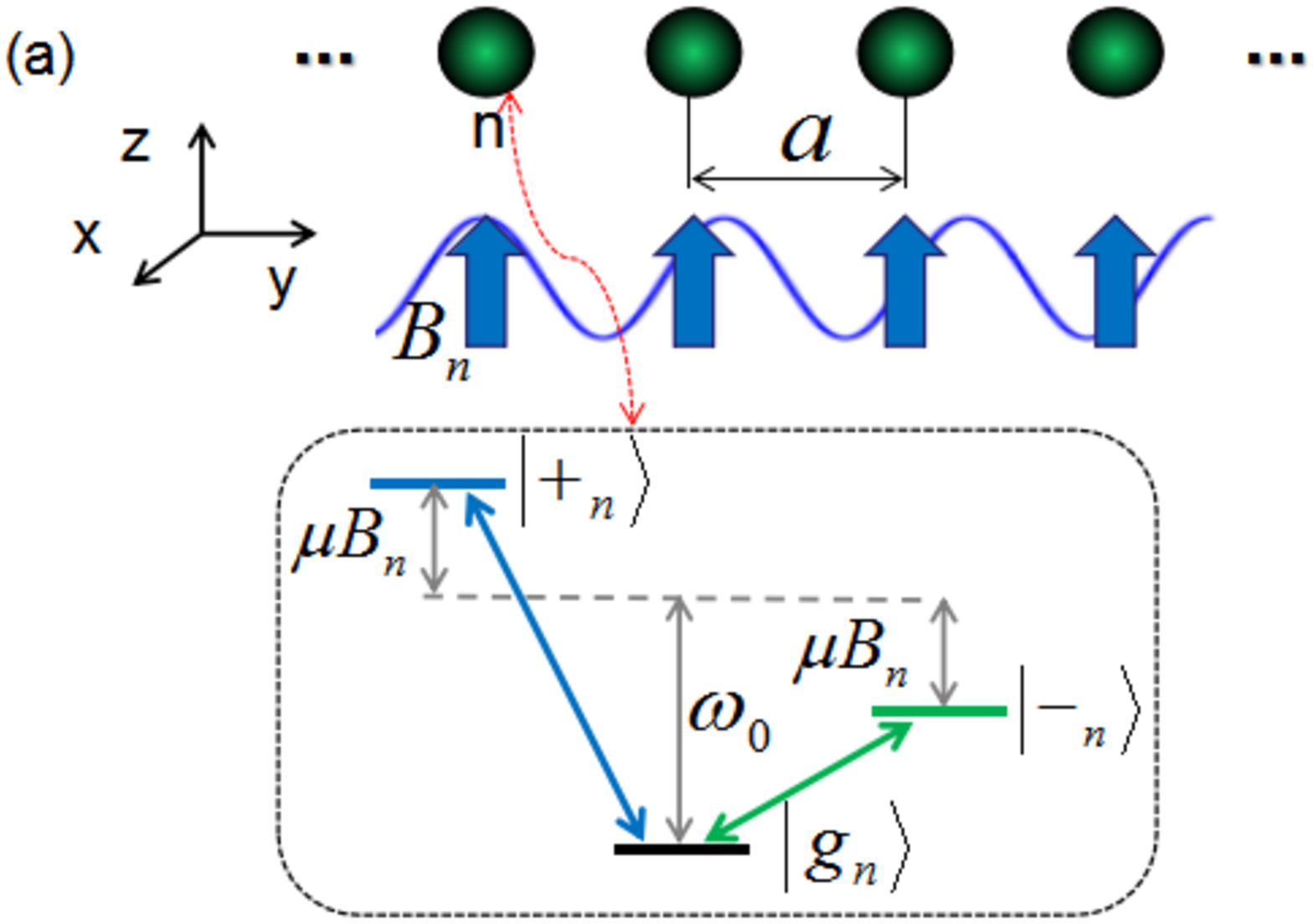}
\includegraphics[width=0.5\textwidth ]{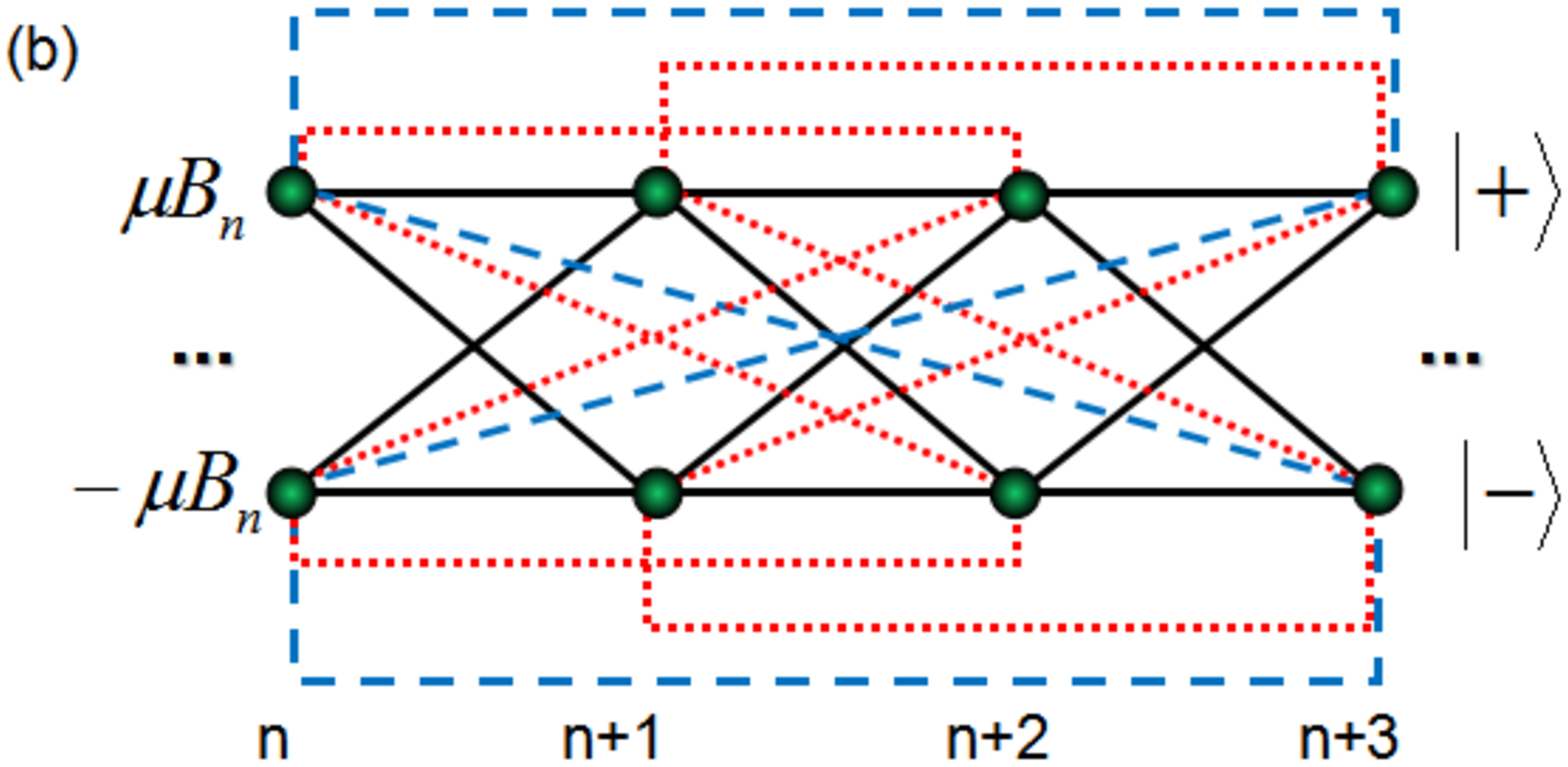}
\caption{(a) An $1$D atomic arrays subjected to an external spatially
periodic magnetic field $B_n$. Each atom has a $V$-type atomic level structure with non-degeneracy excited states $|\pm_n\rangle$ split by the magnetic field. (b) The equivalent tight-binding lattice model with on-site potentials $\pm B_n$ and photon-mediated long-range hoppings. Black solid, red dotted and blue dashed lines label the nearest-neighbor, next-nearest-neighbor and triatomic spacing hoppings, respectively, indicating dipole-dipole interactions and collective dissipations.}\label{figure.1}
\end{figure}

We propose the experimental arrangement consisting of a $1$D arrays of $N$ atoms which are aligned along the $y$ direction with the spacing $a$. Each atom (labelled by $n$ and located at $y_n$) has a $V$-type internal level structure with the ground state $|g_n\rangle$ and excited states $|\pm_n\rangle=\mp(|x_n\rangle\pm i|y_n\rangle)/\sqrt{2}$, where the transition between $|g_n>$ and $|\pm_n>$ is coupled with the right (left) circularly polarized light. Here $|x(y)\rangle$ refers to the state polarized along the $x(y)$ direction.  The degeneracy of the excited states is broken by the presence of an external magnetic field $B_n\equiv B(y_n)$ along the $z$ axis [see Fig.~\ref{figure.1}(a)].

We consider the dynamics of single-excited atoms coupled to free-space modes of the radiation field. After integrating out
 radiation modes under the dipole approximation, one obtains  the non-Hermitian effective Hamiltonian \cite{s0,l2,l3}
\begin{eqnarray}\label{1}
 H&=&\sum^{N}_n \sum_{\alpha=\pm}\bigg(\omega_0-i\frac{\gamma_0}{2}+sgn(\alpha)\mu B_n\bigg)|\alpha_n\rangle\langle\alpha_n|\nonumber \\
 &&+\frac{3\pi\gamma_0}{k_0}\sum^{N}_{n\neq m} \sum_{\alpha,\beta=\pm}G_{\alpha\beta}(y_n-y_m)|\alpha_n\rangle\langle\beta_m|,
\end{eqnarray}
where $\omega_0=k_0c=2\pi c/\lambda$ is the atomic transition frequency with the wave vector $k_0$ and the wavelength $\lambda$, $\gamma_0$ is the atomic decay rate in the free space, $sgn(\pm)\equiv\pm$,  and $\mu B_n$ gives the Zeeman shift  for the $n$th atom with the magnetic moment $\mu$. $G_{\alpha\beta}(y_n-y_m)$ is the free-space dyadic Green$^{,}$s function describing the electric field at $y_n$ emitted by
the atom located at $y_m$.  By using the Green$^{,}$s function in Cartesian basis \cite{l3,g1,s3,g3}, one has
\begin{eqnarray}\label{2}
  G_{\pm\pm} &=& \frac{G_{xx}+G_{yy}}{2}=-\frac{e^{ik_0r}}{8\pi k_0^{2} r^{3}}(k^{2}_0r^{2}-ik_0r+1), \nonumber \\
  G_{\pm\mp} &=& \frac{G_{yy}-G_{xx}}{2}=\frac{e^{ik_0r}}{8\pi k_0^{2} r^{3}}(k^{2}_0r^{2}+3ik_0r-3),
\end{eqnarray}
where $r=|y_n-y_m|$.

The atomic system under investigation is an effective tight-binding lattice model [see Fig.~\ref{figure.1}(b)]. The photon-mediated long-range hoppings amplitude is described by coefficients in the last term of the Hamiltonian, where the real part describes photon-mediated dipole-dipole interaction potential between the $n$th and $m$th atoms, while the imaginary part denotes the collective dissipative rate of the two atoms.

To construct the bandstructure of the non-Hermitian Hamiltonian in Eq.~(\ref{1}), we take the linear
combination of single-excited states $|\psi\rangle=\sum_n (C_{n,+}|+_n\rangle+C_{n,-}|-_n\rangle)$, where $C_{n,\pm}$
is the amplitude of the wave function for the $n$th atom with the $\pm$ polarization. The bandstructure can be calculated by using the time-independent Schr\"{o}dinger equation $H|\psi\rangle=E|\psi\rangle$, which leads to
\begin{eqnarray}\label{3}
  EC_{n,+} &=& \frac{3\pi\gamma_0}{k_0}\sum_{l\neq0}[G_{++}(la)C_{n+l,+}+G_{+-}(la)C_{n+l,-}] \nonumber\\&&+(\omega_0-i\frac{\gamma_0}{2}+\mu B_n)C_{n,+},\nonumber\\
  EC_{n,-} &=& \frac{3\pi\gamma_0}{k_0}\sum_{l\neq0}[G_{-+}(la)C_{n+l,+}+G_{--}(la)C_{n+l,-}]\nonumber\\&&+(\omega_0-i\frac{\gamma_0}{2}-\mu B_n)C_{n,-},
\end{eqnarray}
where $l$ is a nonzero integer. $E\equiv\omega-i\gamma/2$ is the complex eigenvalue, in which $\omega$ denotes the self-energy of the collective atomic excitation and $\gamma$
is the collective decay rate of the system.

We consider a spatially periodic magnetic field
\begin{equation}\label{4}
B_n = B(y_n) =B_0\cos(2\pi bn+\phi),
\end{equation}
where $B_0$ is the amplitude of the magnetic field, $1/b$ is the spatial period,  and $\phi$ is the modulation phase. By applying the magnetic field with different spatial shapes along the $y$ direction, one has the control of parameters $b$ and $\phi$.  Here $\phi$ provides an additional degree of freedom to our system serving the purpose of the synthetic dimension,  so the system can be explored by exploiting the parameter-dependency of  the Hamiltonian \cite{o1,o02}. In such a synthetic space, $b$ gives the effective magnetic flux while $\phi$ denotes a synthetic momentum dimension (reciprocal to a virtual spatial dimension) \cite{o2,yf}. Hence we can study the physics associated to an open $2$D system with long-range couplings under the effective magnetic flux.

 \begin{figure}[htbp]
\centering
\includegraphics[width=0.6\textwidth ]{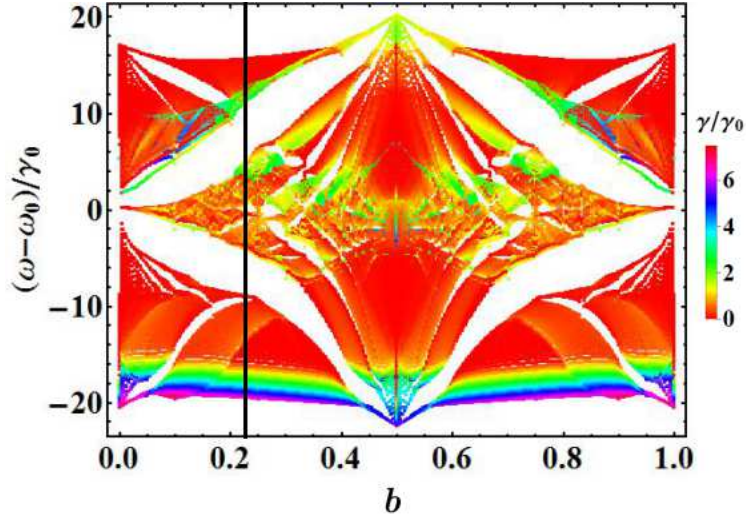}
\caption{The numerically calculated projected bandstructure versus $b$ for the infinite $1$D atomic arrays with $a=0.1\lambda$ and $\mu B_0=10\gamma_0$. The color of the bandstructure gives the collective decay rate $\gamma$.}\label{figure.2}
\end{figure}

We plot the projected bandstructure of the system while varying $b$ in Fig.~\ref{figure.2}. The atomic arrays are assumed to be infinitely long with
$a=0.1\lambda$ and $\mu B_0=10\gamma_0$ \cite{l2}, and bandstructure is computed by following the method in Ref.~\cite{but1}. One can see a butterfly-like bandstructure ($\omega$), which exhibits multiple bulk bands and gaps for each $b$. The bandstructure shows several distinguished features as compared to the Hofstadter butterfly bandstructure in Ref.~\cite{but1} and also the butterfly-like spectrum in the 1D photonic model \cite{new1} due to the long-range non-Hermitian couplings in the atomic arrays \cite{ma}.  The striking feature of our system, as it is seen in Fig.~\ref{figure.2}, is that the collective decay rate ($\gamma$) is changing for different $\omega$ at a certain $b$, covering the range from $0$ to $\sim7.5\gamma_0$.  The destructive interference in the atom-photon interaction leads to suppressed radiative loss for certain bulk states as shown in Fig.~\ref{figure.2}, corresponding to  subradiant states with decay rate below the single-atom emission rate $\gamma_0$. One can therefore control the decay rate to stabilize the quantum coherence in this quantum system.

 \begin{figure}[htbp]
\centering
\includegraphics[width=0.6\textwidth ]{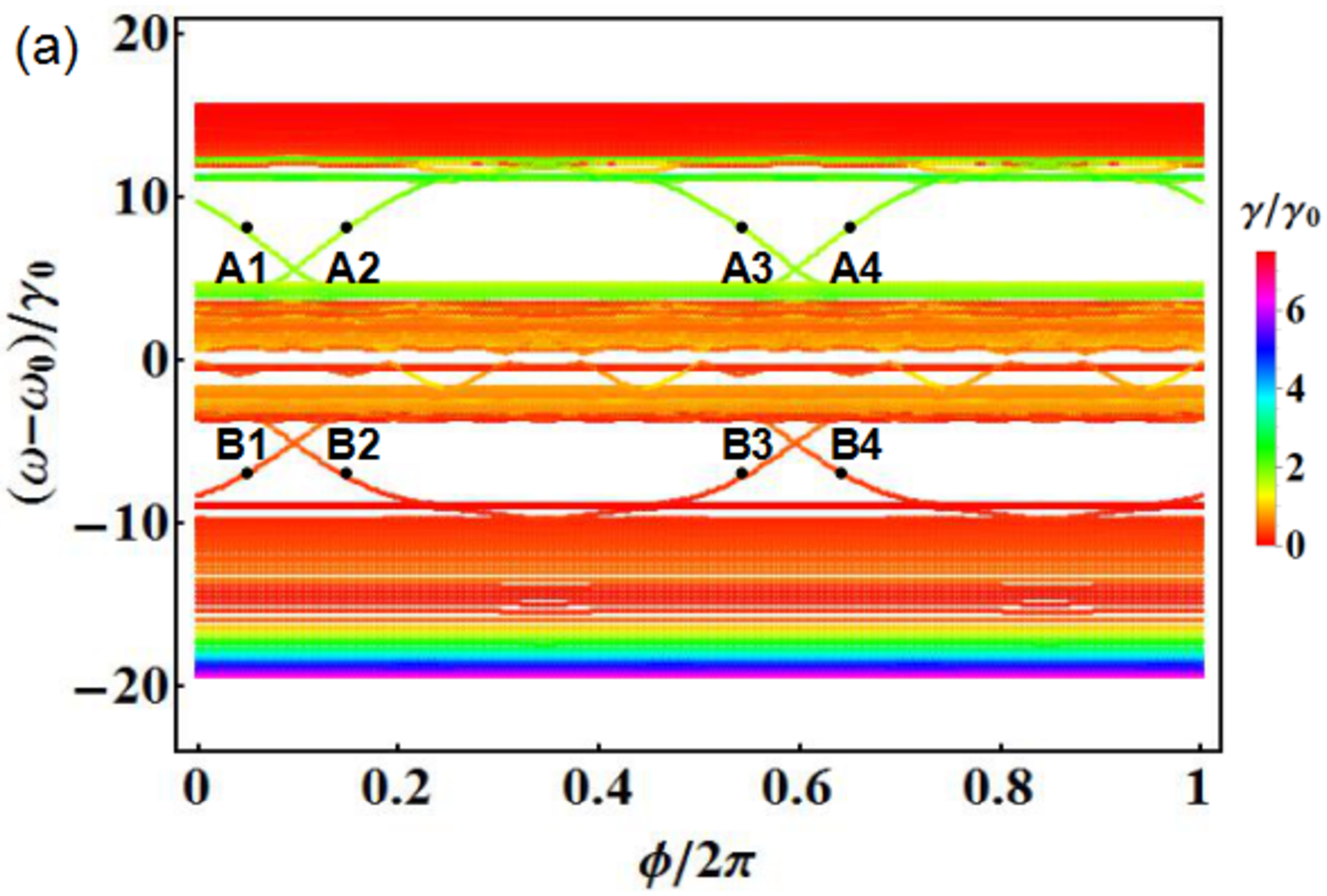}
\includegraphics[width=0.6\textwidth ]{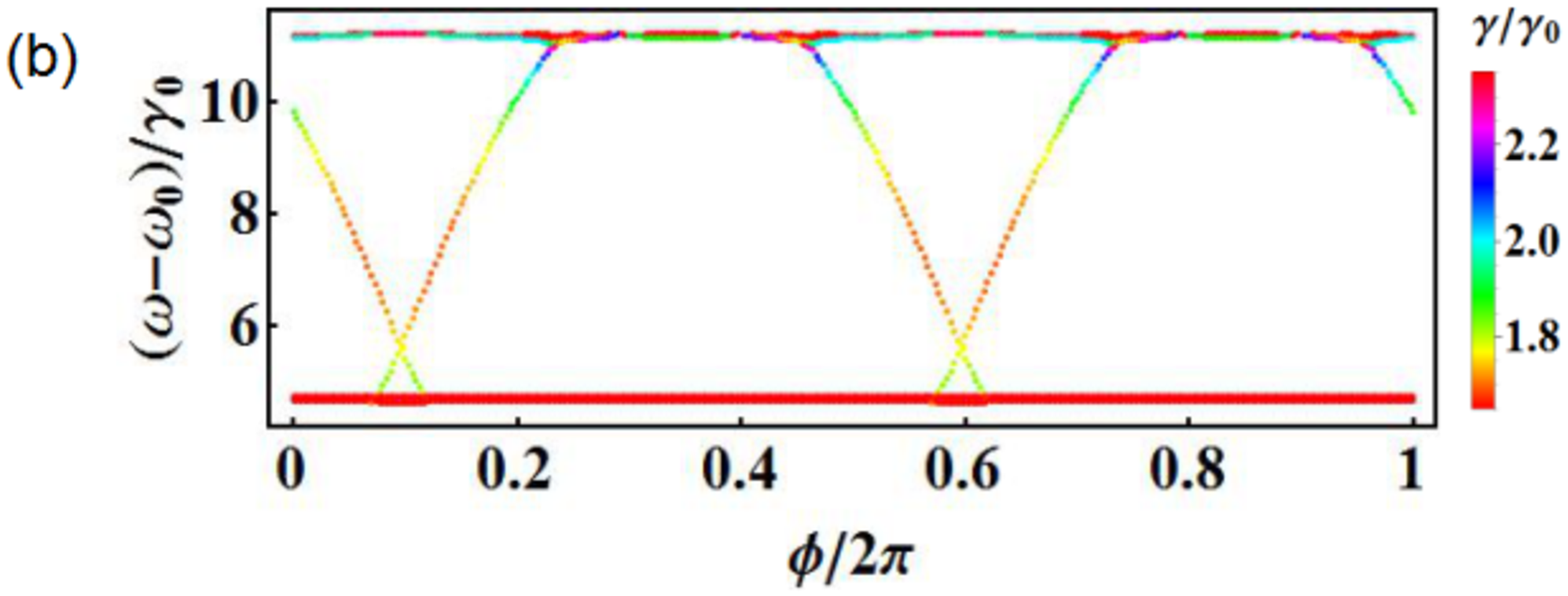}
\includegraphics[width=0.6\textwidth ]{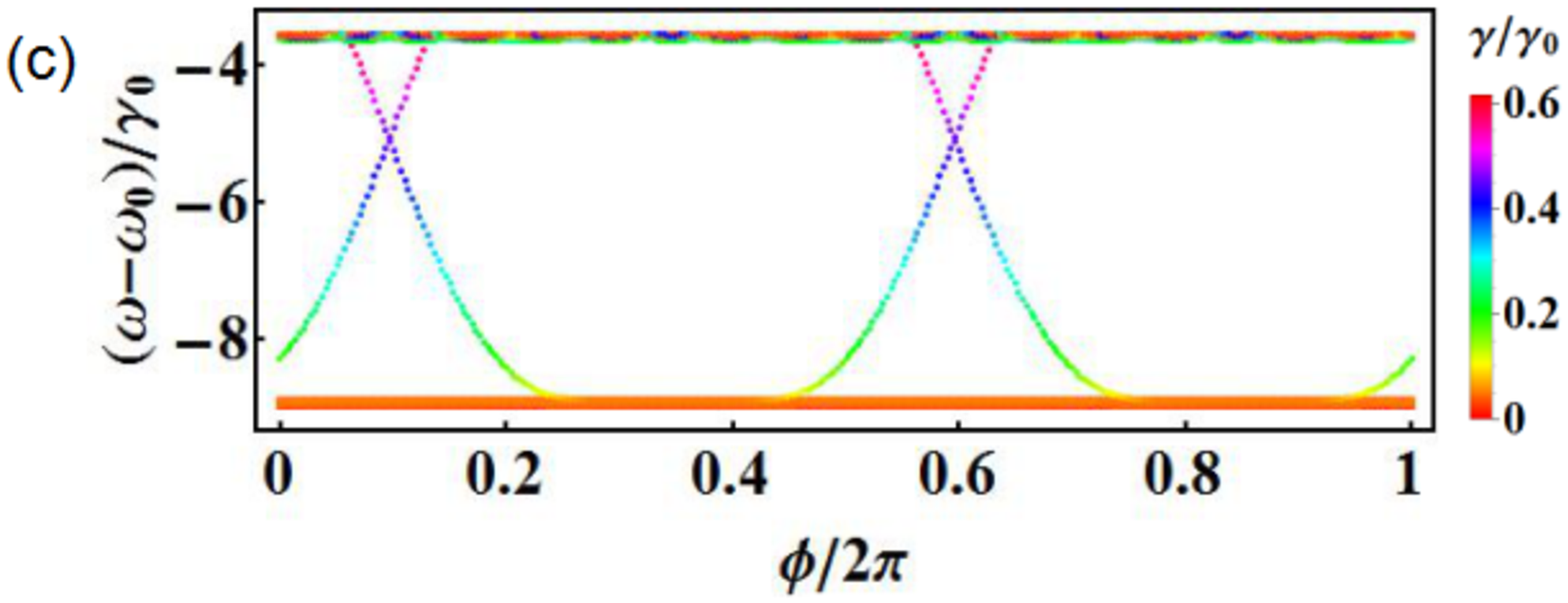}
\caption{(a) The bandstructure as a function of  $\phi$ under open boundary condition for atomic number $N=101$, $a=0.1\lambda$, $\mu B_0=10\gamma_0$ and $b=\sqrt{5}/10$ which corresponds the black line in Fig.~\ref{figure.2}. (b) and (c) The zoom-in bandstructures from (a), showing the details of boundary states. The color of the bandstructures gives the collective decay rate $\gamma$.}\label{figure.3}
\end{figure}
The parameter $b$ can be externally adjusted by controlling the magnetic field. Once it is irrational, the effective magnetic flux is incommensurate with the lattice and the system exhibits a quasicrystal structure \cite{o2}. We set $b=\sqrt{5}/10$, indicated by the black line in Fig.~\ref{figure.2}.
In Fig.~\ref{figure.3}(a),  the bandstructure of the lattice is plotted under an open boundary condition with $N=101$ against the modulation phase $\phi$ in the external magnetic field.
As a consequence of the presence of the magnetic field which breaks the time-reversal symmetry of the system, the bandstructure  is topologically nontrivial.
One can see that there is a fractal set of band gaps, and, inside each gap, it exhibits pairs of topologically protected boundary states. The collective decay rates $\gamma$ for boundary states in two larger gaps show different physical features.
 The boundary states inside the upper gap has $\gamma$ greater than $\gamma_0$, corresponding to superradiant
   modes with enhanced collective emission, while the boundary states inside the lower gap are subradiant because $\gamma$ is smaller than $\gamma_0$. Moreover, the collective decay rate is also changing along each boundary state  when one varies the parameter $\phi$, as shown in Fig.~\ref{figure.3}(b) and~\ref{figure.3}(c). In particular, for the subradiant boundary states in the lower gap, as indicated in Fig.~\ref{figure.3} (c), $\gamma$ changes from $\sim0.6\gamma_0$ to $\sim0.1\gamma_0$, showing a significant suppression of the spontaneous emission. Furthermore, the lifetimes of boundary states are influenced by the choice of the parameter $b$. For instance, for the case $b=(\sqrt{3}-1)/2$,
the boundary states inside the two large gaps are both subradiant as we discuss it in greater details in Supplementary materials \cite{ma}.

 \begin{figure}[htbp]
\centering
\includegraphics[width=0.9\textwidth ]{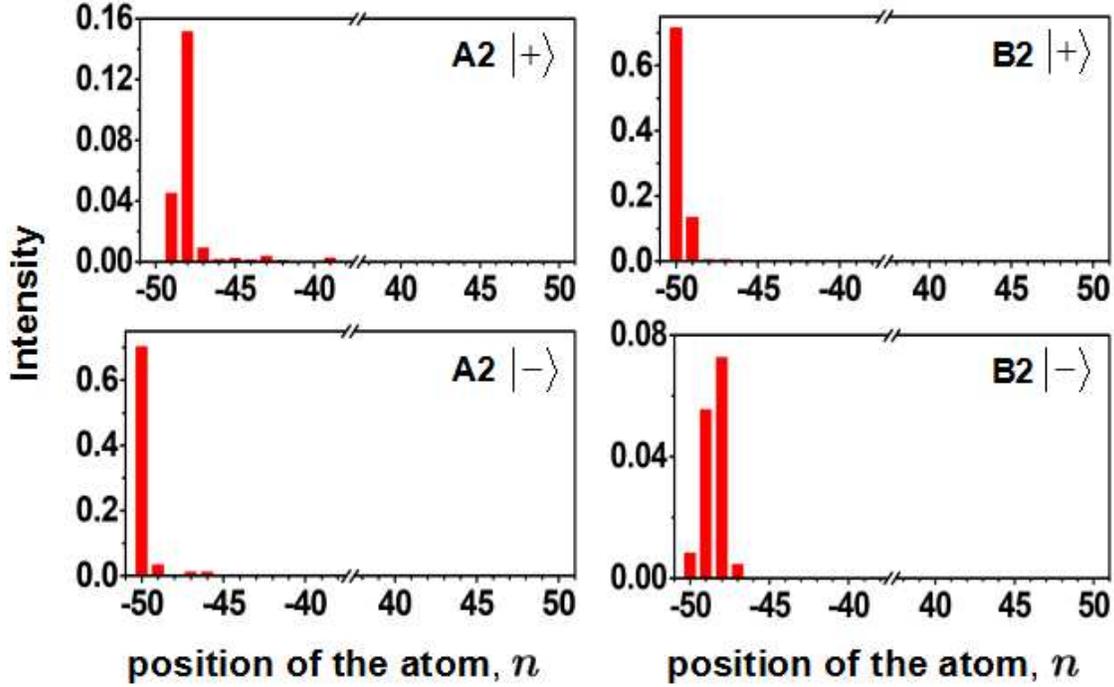}
\caption{Intensity distributions of boundary states with $|+\rangle$ (upper row) and $|-\rangle$ (lower row) excited states, labelled by A2 and B2  in Fig.~\ref{figure.3}(a). Both boundary states are localized on the left boundary of the atomic arrays. A2 and B2 are denoted at $[\phi/2\pi$,$(\omega-\omega_0)/\gamma_0]$=$(0.15,8.15)$ and $(0.15,-7.21)$, respectively.}\label{figure.4}
\end{figure}
The aforementioned boundary states are localized on the left or right boundary of the lattice with a combination of $|+\rangle$ and $|-\rangle$ excited states.
As an example, in Fig.~\ref{figure.4}, we plot the intensity distributions of boundary states versus the position of the atom $n$ for  $|\pm\rangle$ excited states labelled by A2 and B2 in Fig.~\ref{figure.3}(a), which corresponds to superradiant and subradiant  states, respectively. The superradiant boundary state A2 is located mainly at the $|-\rangle$ excited state on the leftmost atom, while a small portion of the intensity is distributed at $|+\rangle$ on other atoms near the left boundary due to the hoppings between the two excited states on different atoms. We denote the boundary state A2 by $L-$ then. Similarly, the subradiant boundary state B2 is located mainly at the $|+\rangle$ excited state on the leftmost boundary, and hence is labelled by $L+$.
The two boundary states corresponding to the same $\phi$ but different excitation frequency, so one can selectively excite either the superradiant or subradiant boundary states for a given external magnetic fields. Such selectively prepared subradiant state localized at the boundary of the arrays, is robust against small variations of the system due to the topological properties. It therefore shows a potential for the robust quantum storage, which is of great importance for quantum device applications.

 Other boundary states labelled by A1, A3, A4 in the upper gap and B1, B3, B4 in the lower gap, as shown in Fig.~\ref{figure.3}(a), give $R_-$ (mainly distributed at $|-\rangle$ on the right  boundary), $R_+$ (mainly distributed at $|+\rangle$ on the right boundary), $L_+$ and $R_+$, $R_-$, $L_-$, respectively.
One therefore can selectively prepare a super- or subradiant state with either right or left circularly polarization by a choice of $\phi$. In each gap, the boundary states located at the same boundary with opposite polarization excitations exhibit propagating modes toward the same direction along the virtual spatial dimension (reciprocal to the synthetic momentum dimension $\phi$), while the boundary states at different boundaries support propagating modes towards opposite direction.

The proposed system  is experimentally feasible. For example, an atomic array with the subwavelength-scale lattice spacing can be realized by using bosonic strontium \cite{str,str2}. The transition between triplet states $^{3}P_0$ and $^{3}D_1$ of atom $^{84}$Sr gives emission at the wavelength $\lambda=2.6\mu m$.
One can use the optical lattice formed by lasers at $412.8$nm to trap the atoms, which achieves a subwavelength lattice spacing $a=206.4$nm, i.e., $a/\lambda\approx  0.08$ \cite{str}.
Inhomogeneous magnetic field is widely used to produce spin-orbit couplings in the condensed matter systems \cite{th0,th}. The magnetic field in Eq.~(\ref{4}) can be implemented by a variety of experimental technologies which have been  proposed to construct magnetic lattices \cite{new2,an13,new3,new4,new5}. One can use a circularly polarized laser at a frequency resonant with the boundary state $L_{\pm}$ ($R_{\pm}$) inside the band gap to excite the $|\pm\rangle$ state of the atom located at the left (right) boundary. The emission of such  super- or subradiant boundary state is localized at the boundary of atomic arrays with the enhanced or suppressed collective decay rate.

In summary, we have investigated $1$D atomic arrays subjected to
the spatially periodic magnetic field, which supports the
non-Hermitian lattice model with long-range interactions. The
phase in the magnetic field serves as an external parameter, which
gives a synthetic momentum dimension. In atomic arrays with strong
long-range interactions, the bulk-boundary correspondence is not
generally valid \cite{l4}. In the open system proposed here, we
consider a synthetic space including one spatial dimension and one
synthetic momentum dimension. By carefully selecting parameters,
we show the existence of the $2$D bulk-boundary correspondence in
this synthetic space that exhibits topologically-protected
boundary states, which holds fundamentally
different physics from the 1D atomic arrays with a non-zero Zak
phase \cite{g3}. These boundary states are localized at the
boundary of atomic arrays and exhibit topologically-protected
super- or subradiance with right or left circularly polarization.
Our results show  potential applications towards manipulating
atomic emission at the single photon level under the topological
protection, which is important for the quantum
memory and quantum information, and also leads potential
implications for quantum sensors \cite{impact1,impact2} and
super-resolution spectroscopy \cite{impact3} by using the robust
single-photon superradiant states. The study of topological
quantum optics with synthetic dimensions opens a route of
exploring topological phenomena in versatile higher-dimensional
strong-interacting open quantum systems.

\begin{acknowledgments}
This paper is supported by the National Nature Science Foundation of China (NSFC) No. 11734011 and the National Key R\&D Program of China (2017YFA0303701).
\end{acknowledgments}





\begin{thebibliography}{81}
\bibitem{ww} W. E. Lamb, J. and R. C. Retherford, Phys. Rev. {\bf72}, 241 (1947).
\bibitem{di1} R. Dicke, Phys. Rev. {\bf93}, 99 (1954).




\bibitem{2} T. Xia, M. Lichtman, K. Maller, A. W. Carr, M. J.
Piotrowicz, L. Isenhower, and M. Saffman, Phys. Rev. Lett.
{\bf114}, 100503 (2015).
\bibitem{3} M. Endres, H. Bernien, A. Keesling, H. Levine, E. R.
Anschuetz, A. Krajenbrink, C. Senko, V. Vuletic, M.
Greiner, and M. D. Lukin, Science {\bf354}, 1024 (2016).

\bibitem{6} D. Barredo, S. de L\'{e}s\'{e}leuc, V. Lienhard, T. Lahaye, and A.
Browaeys, Science {\bf354}, 1021 (2016).





 \bibitem{s1}  G. Facchinetti, S. D. Jenkins, and J. Ruostekoski, Phys. Rev.
Lett. {\bf117}, 243601 (2016).

 \bibitem{s3} A. Asenjo-Garcia, M. Moreno-Cardoner, A. Albrecht,
H. J. Kimble, and D. E. Chang, Phys. Rev. X {\bf7}, 031024 (2017).


\bibitem{s51}P.-O. Guimond, A. Grankin, D. V. Vasilyev, B. Vermersch, and P. Zoller,
Phys. Rev. Lett. {\bf122}, 093601 (2019).
\bibitem{s6} F. J. Garc\'{i}a De Abajo, Rev. Mod. Phys. {\bf79}, 1267 (2007).

\bibitem{s8}R. J. Bettles, S. A. Gardiner, and C. S. Adams, Phys. Rev.
Lett. {\bf116}, 103602 (2016).

\bibitem{s0} E. Shahmoon, D. S. Wild, M. D. Lukin, and S. F. Yelin,
Phys. Rev. Lett. {\bf118}, 113601 (2017).
\bibitem{m1} M. T. Manzoni, M. Moreno-Cardoner, A. Asenjo-Garcia,
J. V. Porto, A. V. Gorshkov, and D. E. Chang, New J. Phys.
{\bf20}, 083048 (2018).

\bibitem{l2}
J. Perczel, J. Borregaard, D. E. Chang, H. Pichler, S. F. Yelin, P. Zoller, and M. D. Lukin, Phys. Rev. Lett. {\bf119}, 023603 (2017).
\bibitem{l4} R. J. Bettles, J. Min\'{a}\v{r}, C. S. Adams, I. Lesanovsky, and B. Olmos, Phys. Rev. A {\bf96}, 041603(R) (2017).
\bibitem{k1} K. Hammerer, A. S. S{\o}rensen, and E. S. Polzik,  Rev. Mod. Phys. {\bf82}, 1041 (2010).

\bibitem{ef1} K. V. Klitzing, G. Dorda, and M. Pepper, Phys. Rev. Lett. $\textbf{45}$, 494 (1980).
\bibitem{ef2} D. C. Tsui, H. L. Stormer, and A. C. Gossard, Phys. Rev. Lett. $\textbf{48}$, 1559 (1982).
\bibitem{ef3} R. B. Laughlin, Phys. Rev. Lett. $\textbf{50}$, 1395 (1983).

\bibitem{tp3} T. Ozawa, H. M. Price, A. Amo, N. Goldman, M. Hafezi, L. Lu, M. C. Rechtsman, D. Schuster, J. Simon, O. Zilberberg, and I. Carusotto,
 Rev. Mod. Phys. $\textbf{91}$, 015006 (2019).
 \bibitem{gg4} N. R. Cooper, J. Dalibard, and I. B. Spielman, Rev. Mod. Phys. $\textbf{91}$, 015005 (2019).
\bibitem{gg5} D.-W. Zhang, Y.-Q. Zhu, Y. X. Zhao, H. Yan, and S.-L. Zhu, Advances in Physics, $\textbf{67}$, 253 (2019).


\bibitem{h1} Z. Wang, Y. Chong, J. D. Joannopoulos, and M. Solja\v{c}i\'{c}, Nature $\textbf{461}$, 772 (2009).
\bibitem{h2}M. Hafezi, E. Demler, M. Lukin, J. Taylor, Nat. Phys. $\textbf{7}$, 907 (2011).
\bibitem{h3} K. Fang, Z. Yu, S. Fan, Nat. Photon. $\textbf{6}$, 782 (2012).
\bibitem{h4} M. C. Rechtsman, J. M. Zeuner, Y. Plotnik, Y. Lumer, D. Podolsky, F. Dreisow, S. Nolte, M. Segev, and A. Szameit, Nature $\textbf{496}$, 196 (2013).



\bibitem{dw1}D.-W. Wang, H. Cai, L. Yan, S.-Y. Zhu, R.-B. Liu, Optica $\textbf{2}$, 712 (2015).
\bibitem{dw2} H. Cai, J. Liu, J. Wu, Y. He, S.-Y. Zhu, J.-X. Zhang, D.-W. Wang, Phys. Rev. Lett. $\textbf{122}$, 023601 (2019).

\bibitem{l3}
J. Perczel, J. Borregaard, D. E. Chang, H. Pichler, S. F. Yelin, P. Zoller, and M. D. Lukin, Phys. Rev. A {\bf96}, 063801 (2017).
\bibitem{g3}
B. X. Wang and C. Y. Zhao,  Phys. Rev. A {\bf98}, 023808 (2018).
\bibitem{gg3} J. Perczel, J. Borregaard, D. E. Chang, S. F. Yelin, M. D. Lukin, arxiv:1810.12299 (2018).




\bibitem{o1} L. Yuan, Q. Lin, M. Xiao, and S. Fan, Optica $\textbf{5}$, 1396 (2018).
\bibitem{o02}T. Ozawa and H. M. Price, Nat. Rev. Phys. (2019). DOI: 10.1038/s42254-019-0045-3.

\bibitem{im3}M. Lohse, C. Schweizer, H. M. Price, O. Zilberberg, and I. Bloch, Nature $\textbf{553}$,
55 (2018).

\bibitem{im6} O. Zilberberg, S. Huang, J. Guglielmon, M. Wang, K. P. Chen, Y. E. Kraus, and M. C. Rechtsman,  Nature $\textbf{553}$,
59 (2018).



\bibitem{im2} E. Lustig, S. Weimann, Y. Plotnik, Y. Lumer, M. A. Bandres, A. Szameit, and M. Segev, Nature $\textbf{567}$, 356 (2019).

\bibitem{im8} A. Dutt, M. Minkov, Q. Lin, L. Yuan, D. A. B. Miller, and S. Fan, arxiv.org/abs/1903.07842.

\bibitem{new1} L. Lang, X. Cai, and S. Chen, Phys. Rev. Lett. \textbf{108}, 220401 (2012).
\bibitem{o2} Y. E. Kraus, Y. Lahini, Z. Ringel, M. Verbin, and O. Zilberberg, Phys. Rev. Lett. $\textbf{109}$, 106402 (2012).

\bibitem{g1}
O. Morice, Y. Castin, and J. Dalibard, Phys. Rev. A {\bf51}, 3896 (1995).


\bibitem{yf} C. Shang, X. Chen, W. Luo, and F. Ye, Opt. Lett. $\textbf{43}$, 275 (2018).

\bibitem{but1}
D. R. Hofstadter, Phys. Rev. B {\bf14}, 2239 (1976).
\bibitem{ma} See the Supplemental Materials.


\bibitem{str}B. Olmos, D. Yu, Y. Singh, F. Schreck, K. Bongs, and I. Lesanovsky, Phys. Rev. Lett. $\textbf{110}$, 143602 (2013).
\bibitem{str2}S. V. Syzranov, M. L. Wall, V. Gurarie, and A. M. Rey, Nat. Commun. $\textbf{7}$, 13543 (2016).
\bibitem{th0}Y. Tokura, W. G. van der Wiel, T. Obata, and S. Tarucha, Phys. Rev. Lett. $\textbf{96}$, 047202 (2006).
\bibitem{th} M. Pioro-Ladri\`{e}re, T. Obata, Y. Tokura, Y.-S. Shin, T. Kubo, K. Yoshida, T. Taniyama, and S. Tarucha,
 Nat. Phys. $\textbf{4}$, 776 (2008).

\bibitem{new2} A. D. West, K. J. Weatherill, T. J. Hayward, P. W. Fry, T. Schrefl, M. R. J. Gibbs, C. S. Adams, D. A. Allwood, and I. G. Hughes, Nano Lett. \textbf{12}, 4065 (2012).
\bibitem{an13} B. M. Anderson, I. B. Spielman, and G. Juzeli\={u}nas, Phys. Rev. Lett. $\textbf{111}$, 125301 (2013).
\bibitem{new3} X. Luo, L. Wu, J. Chen, R. Lu, R. Wang, and L. You, New J. Phys. \textbf{17}, 083048 (2015).
\bibitem{new4} Y. Wang, P. Surendran, S. Jose, T. Tran, I. Herrera, S. Whitlock, R. McLean, A. Sidorov, and P. Hannaford, Sci. Bull. \textbf{61}, 1097 (2016).
\bibitem{new5} J. Yu, Z. Xu, R. L\"{u}, and L. You, Phys. Rev. Lett. \textbf{116}, 143003 (2016).

\bibitem{impact1} J. Kitching, S. Knappe, and E. A. Donley, IEEE Sens. J. \textbf{11}, 1749 (2011).
\bibitem{impact2} S. J. Roof, K. J. Kemp, and M. D. Havey, Phys. Rev. Lett. \textbf{117}, 073003 (2016).
\bibitem{impact3} A. von Diezmann, Y. Shechtman, and W. E. Moerner, Chem. Rev. \textbf{117} 7244 (2017).
\end{thebibliography}

\end{document}